\documentclass[10pt, conference, letterpaper]{IEEEtran}
\IEEEoverridecommandlockouts
\usepackage{cite}
\usepackage{amsmath,amssymb,amsfonts}
\usepackage{algorithmic}
\usepackage{graphicx}
\usepackage{textcomp}
\usepackage{xcolor}
\usepackage{arydshln}
\usepackage{ulem}
\usepackage{multirow}

\def\BibTeX{{\rm B\kern-.05em{\sc i\kern-.025em b}\kern-.08em
    T\kern-.1667em\lower.7ex\hbox{E}\kern-.125emX}}
\begin{document}

\title{
Learning-Based WiFi Fingerprint Inpainting via Generative Adversarial Networks
}

\author{\IEEEauthorblockN{ Yu Chan, Pin-Yu Lin, Yu-Yun Tseng,  Jen-Jee Chen, and Yu-Chee Tseng}
\IEEEauthorblockA{\textit{College of Artificial Intelligence} \\
\textit{National Yang Ming Chiao Tung University}\\
Tainan City, Taiwan \\
charlie88162@gmail.com, qaz1232009.ai09@nycu.edu.tw, yuts1923@colorado.edu, \\ jenjee@nycu.edu.tw, yctseng@nycu.edu.tw}
}

\maketitle

\begin{abstract}
WiFi-based indoor positioning has been extensively studied. A fundamental issue in such solutions is the collection of WiFi fingerprints. However, due to real-world constraints, collecting complete fingerprints at all intended locations is sometimes prohibited. This work considers the WiFi fingerprint inpainting problem. This problem differs from typical image/video inpainting problems in several aspects. Unlike RGB images, WiFi field maps come in any shape, and signal data may follow certain distributions. Therefore, it is difficult to forcefully fit them into a fixed-dimensional matrix, as done with processing images in RGB format. As soon as a map is changed, it also becomes difficult to adapt it to the same model due to scale issues. Furthermore, such models are significantly constrained in situations requiring outward inpainting. Fortunately, the spatial relationships of WiFi signals and the rich information provided among channels offer ample opportunities for this generative model to accomplish inpainting. Therefore, we designed this model to not only retain the characteristic of regression models in generating fingerprints of arbitrary shapes but also to accommodate the observational outcomes from densely deployed APs. This work makes two major contributions. Firstly, we delineate the distinctions between this problem and image inpainting, highlighting potential avenues for research. Secondly, we introduce novel generative inpainting models aimed at capturing both inter-AP and intra-AP correlations while preserving latent information. Additionally, we incorporate a specially designed adversarial discriminator to enhance the quality of inpainting outcomes.
\end{abstract}

\begin{IEEEkeywords}
adversarial network, indoor localization, inpainting, WiFi fingerprint, wireless sensing
\end{IEEEkeywords}

\section{Introduction}
The popularity of IoT, cloud services, and social networks has promoted the expansion of WiFi terminals. With more WiFi hotspots being deployed, WiFi-based localization may continue to improve with nearly no additional cost~\cite{wifi_survey_1}. Active WiFi positioning can be classified as {\it range-based} and {\it range-free}. Range-based methods utilize Time of Arrival (ToA), Angle of Arrival (AoA), or signal strength to estimate locations, but may suffer from multipath effects. Range-free methods~\cite{wifi_pos_2,wifi_pos_1} match the collected RSS (Received signal strength) with a predefined WiFi fingerprint map and are thus more resilient to multi-path effects~\cite{wifi_survey_3}. Traditional RSS-based solutions are rule-based. Recently, several learning-based solutions have been proposed~\cite{wifi_pos_3,wifi_pos_2,wifi_pos_5,wifi_pos_survey_1}.

A basic step in all RSS-based solutions is fingerprints collection, which is labor-intensive and error-prone. To reduce manual efforts fingerprints collection can be done by automated robots~\cite{robot_collect_1,robot_collect_2,robot_collect_3} or by crowdsourcing~\cite{crowdsourcing_1,crowdsourcing_2,crowdsourcing_3,crowdsourcing_4,crowdsourcing_5}. Several works have addressed how to apply deep learning to better utilize crowdsourced data~\cite{crowdsourcing_6,crowdsourcing_8,crowdsourcing_9}.

On the other hand, fingerprint inpainting is to fill up fingerprints in those unsurveyed regions~\cite{ProxyFAUG,wifi_survey_3,wifi_survey_4,wifi_pos_survey_1,wifi_pos_5,crowdsourcing_4}. These missing data may damage the integrity of a fingerprint map and thus positioning accuracy. Although image and video inpainting has been intensively studied in the literature~\cite{image_inpaint1,image_inpaint2,image_inpaint3,image_inpaint4,video_inpaint1,video_inpaint2,video_inpaint3,MPVF}, fingerprint inpainting poses  multiple challenges as follows:
\begin{itemize}
\item
While images are normally of a rectangle shape, a field map can be of any shape, and so are the unsurveyed regions.
\item
While images have discrete (grid) points, the points to be inpainted may fall in a continuous space. 
\item Every field has its unique partitions, and its WiFi fingerprints highly depend on the physical partitions. Therefore, trying to understand the physical rules governing WiFi signals by studying data from other fields may offer limited assistance. 

\item
WiFi signal propagation follows physical laws, which are different from natural visions. In image world, every pixel has only one value. However, we usually make multiple RSS observations at a survey position.

\item The decay rules of WiFi signals are consistent worldwide. Whether it is possible to leverage WiFi rules learned from other locations will be a key question. 

\end{itemize}

Nevertheless, there is a factor that could greatly assist the fingerprint inpainting task. In modern buildings, WiFi access points are densely deployed, providing numerous channels for exploitation during the inpainting process. For example, in a fingerprint map $M$, there are two data points $(a,(f_{a-{AP1}},f_{a-{AP2}}))$ and $(b,(f_{b-{AP1}},f_{b-{AP2}}))$, where $a$ and $b$ represent two data collection points, and $f_{p-{APx}}$ denotes the measured RSSI strength of $APx$ at point $p$. Even if the distances from points $a$ and $b$ to $AP1$ are quite far, resulting in similar small values for $f_{a-{AP1}}$ and $f_{b-{AP1}}$, thus hindering the conversion of signal strength into spatially relevant features, we can still have the chance to learn the relationship between $f_{a-{AP2}}$ and $f_{b-{AP2}}$ to determine their correlation. This work proposes innovative solutions - a specialized model for adapting to inpainting WiFi data. Besides, we try to preserve spatial relationships as much as possible. We take advantage of deploying a sufficient number of APs (both in quantity and spatial distribution) within a space, which enables us to understand the relative relationships of all positions in that space. It introduces two models capable of learning within a given fingerprint map and directly inpainting on that given fingerprint map. We regard multiple APs as multiple channels and explore both inter- and intra-AP correlations. In essence, this learning process involves fitting local data into the model. For every unique space, such steps are irreplaceable. Through such a design, our models are able to explore both inter- and intra-AP correlations during the training process.

This work makes several contributions. Through inpainting, it makes RSS-based positioning possible even if some regions of the field are not pre-surveyed. In terms of model design, we have created a model that retains the capability of regression to generate fingerprints for arbitrary fields while accommodating the observational outcomes from densely deployed APs. Our experiences show that collecting a fingerprint at a point takes about 2 seconds. So taking, say, 50 samples per point costs about 3 minutes. Our work greatly relieves the survey cost. We also find that counter-intuitively, inpainting an exterior area is not necessarily more difficult than inpainting an interior area, in terms of loss, and that a minimum of 8 APs per position is sufficient to make our models work well. Through deep learning, inpainting the missing fingerprints of a map is possible without relying on complex propagation equations.

The rest of this paper is organized as follows. Section 2 reviews some related works. Section 3 presents our IAP and I2AP models. Our experiment results are in Section 4. Conclusions are drawn in Section 5.

\section{Related Works}
Indoor positioning is an essential issue \cite{mag, aoa-mobisys}. Wireless positioning systems can be categorized into range-based and range-free approaches. A survey and comparative study of fingerprint-based approaches is in~\cite{wifi_pos_alg_1}. Distance estimation is determined by an artificial neural network model in~\cite{wifi_pos_alg_3}. A Fletcher-Reeves conjugate gradient-based multi-layer feedforward neural network is proposed in~\cite{wifi_pos_alg_4}. A comparison of machine learning indoor algorithms is in~\cite{wifi_pos_5}. Autonomous crowdsourcing of handheld devices is discussed in~\cite{crowdsourcing_4}. A new boundary (indoor and outdoor) localization scheme is proposed in~\cite{boundary}. 

Data augmentation can help improve positioning accuracy. In~\cite{Super_resolution}, a grid-wise fingerprint map is considered as an image and a super-resolution technique is applied to enhance the details of the map. Inpainting has also been studied in ~\cite{wifiGAN_inpaint,kim2021access,han2020radio,belmonte2019recurrent}, but these work cannot handle arbitrary fields and arbitrary locations due to their grid constraints. Reference~\cite{GPR-GAN} regards the inpainting problem as a domain transfer problem from GPR (Gaussian process regression) to ground truth data. A proximity-based fingerprint augmentation is proposed in~\cite{ProxyFAUG}. These works~\cite{ProxyFAUG,yang2020semi,yang2020indoor} cannot inpaint data for exterior areas. How to enhance a fingerprint map is discussed in~\cite{kuo2008scrambling,moghtadaiee2019new}. A selective generative model for fingerprint augmentation is proposed in~\cite{wifiGAN_inpaint}. General data imputation has also been intensively studied~\cite{inpainting_1,inpainting_2}, but we will omit the details.

When it comes to WiFi fingerprints, crowdsourcing data is an inevitable choice. How to exploit crowdsourcing is another challenge. The work~\cite{niu2015wicloc} uses crowdsourced data to determine the site survey fingerprint of an interested location. A hidden Markov model and a ratio-based map-matching algorithm are proposed in~\cite{fu2017crowdsourcing} to automatically update fingerprints. Mapping crowdsourced data to a floor plan is addressed in~\cite{wu2014smartphones}. Reference~\cite{zhao2018crowdsourcing} proposes a method to combine crowdsourcing, floor plan, and motion sensor data. It is claimed that crowdsourced fingerprints are more suitable for localizing moving pedestrians than site survey fingerprints~\cite{wei2021efficient}. However, in this paper, we will not focus on the discussion regarding crowdsourced data. Our work is inspired by \cite{GPR-GAN}, which proposes a generative model that preserves spatial relationships as much as possible in the form that best adapts to WiFi data.

\section{WiFi Fingerprint Inpainting}
We consider a 2D field deployed with $n$ WiFi APs. For any point $p$ in the field, its fingerprint $f_p$ is denoted as an $n$-tuple $f_p = (S_1, S_2, \dots, S_n)$, where $S_i, i=1\dots n$, is the sampled RSS of the $i$th AP at $p$. A fingerprint map $M$ is a set of pairs $(p,f_p)$ such that $p$ is a surveyed point and $f_p$ is a sampled fingerprint. Given $M$ and any unsurveyed point $u$ in the field, the WiFi fingerprint inpainting problem is to predict a fingerprint $f_u$ at $u$ with the minimum loss compared to the ground truth. The unsurveyed points may form regions. Note that to conquer the WiFi signal fluctuation problem, a point may be surveyed multiple times. So $M$ is a multi-set and there could be multiple ground truth for $f_u$. Similarly, an inpainting model may generate multiple $f_u$. These inpainted fingerprints are to enrich $M$ for increasing localization accuracy.

We propose two models. The first model begins with using GPR to predict an initial result and then explores inter-AP relationships. In contrast, the second model entirely relies on deep learning to explores both inter- and intra-AP correlations.


\subsection{Inter-AP (IAP) Inpainting Model}
\begin{figure*} [ht]
    \centering
    \includegraphics[width=0.9\textwidth]{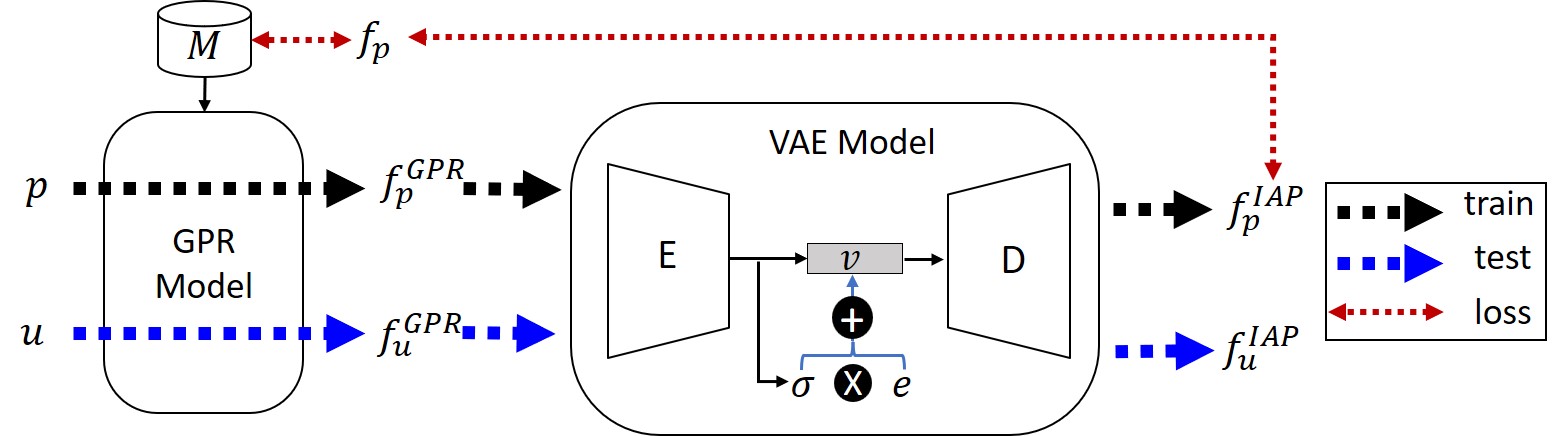}
    \caption{IAP model architecture.}
    \label{fig:iap-arch}
\end{figure*}

This model is inspired by GPR-GAN~\cite{GPR-GAN}. We first use GPR to calculate an initial fingerprint for $u$, and then apply a Variational AutoEncoder (VAE) to improve the result. Figure~\ref{fig:iap-arch} shows our IAP architecture. First, we use $M$ to train a GPR model~\cite{GPR}. Given any point $p$ in $M$, let the output of the GPR model be $f^{GPR}_{p}$. Then the VAE model takes $f^{GPR}_{p}$ to to generate the latent variable $V$, which is then utilized to generate an improved prediction $f^{IAP}_{p}$. The VAE loss function is defined as $\alpha L_{rec} + \beta L_{KL}$, where $L_{rec}$ is the reconstruction loss, $L_{KL}$ is KL-divergence, and $\alpha$ and $\beta$ are hyperparameters: 
\begin{equation}
L_{rec} = \sum_{(p,f_p)\in M}|f^{IAP}_{p}-f_p|
\label{eq:reconstruction}
\end{equation}
\begin{equation}
L_{KL} = D_{KL}(V \| \mathcal{N}(0, 1))
\label{eq:kl}
\end{equation}

The model is trained by the data in $M$. For actual inpainting, we input $u$ into GPR to obtain $f^{GPR}_{u}$, which is forwarded to the VAE model to obtain $f^{IAP}_{u}$. Compared to GPR-GAN, we observe that GAN does not provide adequate supervision on its discriminator's loss. Without strong supervision, GAN is less likely to converge without sufficient training samples, especially in a complex indoor environment. Tuning the hyperparameters of a GAN model stands as another challenge. In~\cite{GPR-GAN}, GPR-GAN is tested on a dataset with 50 samples per position. The loss function of IAP places strong supervision and our model can be well-trained with around 10 samples per point. However, the deficiencies of IAP include:

\begin{description}
\item[1.] VAE highly relies on GPR, which is in turn constrained by AP stability.
\item[2.] The hypothesis of IAP is that VAE conducts domain transformation from GPR prediction to ground truth. However, this domain projection might differ greatly between training points and testing points, making the transformation fail sometimes.
\item[3.] VAE makes predictions without relying on the location of $u$. That is, spatial information is provided to GPR, but such information fades away in the middle of VAE. 
\end{description}
These observations motivate us to design the next model.

\subsection{Inter- and Intra-AP (I2AP) Inpainting Model}

Different from IAP, the I2AP model no longer relies on an external mathematical tool. It explores the spatial relationships of both inter- and intra-AP fingerprint patterns and directly predicts a fingerprint for $u$. Multiple APs in $M$ can be regarded as a gradient image with multiple channels. In addition to fingerprint generation, a specialized discriminator is designed to enhance the prediction result. 

\begin{figure*}
    \centering
    \includegraphics[width=0.9\textwidth]{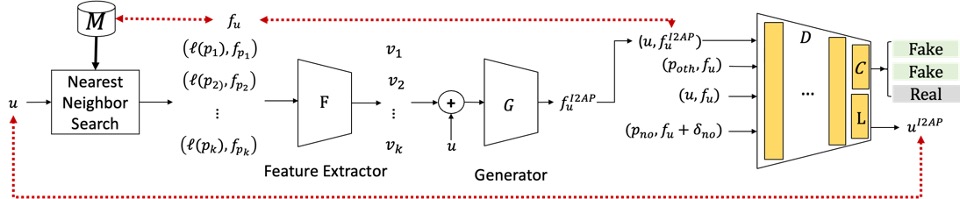}
    \caption{I2AP model architecture.}
    \label{fig:I2AP-arch}
\end{figure*}

Figure~\ref{fig:I2AP-arch} shows the I2AP architecture. Given point $u$, the model first searches for $k$ nearest neighbors of $u$ from $M$, where $k$ is pre-defined. Let these $k$ points be $p_1,p_2,\dots,p_k$ and their fingerprints be $f_{p_1},f_{p_2},\dots,f_{p_k}$, respectively. (Note that since $M$ is a multi-set, if a point has multiple fingerprints in $M$, we can randomly sample one.) For each $p_i,i=1,\dots,k$, we construct an $(n+2)$-tensor $( \ell(p_i),f_{p_i})$, where $\ell(p_i)$ is the (2D) location of $p_i$. The feature extractor $F$ then transforms each tensor $(\ell(p_i),f_{p_i})$ to a feature vector $v_i$. These $v_i$s and $u$ are then concatenated as a 1D $(k(n+2)+2)$-tensor, which is sent to the generator $G$ as input. $G$'s output $f^{I2AP}_u$ is the inpainting result. Two points worth mentioning is that the dimension of $v_i$ should be relatively larger than the fingerprint dimension, $n$. Otherwise, the vector would not be able to retain all information of a fingerprint. Furthermore, the value of $k$ also requires consideration. If it is set to 1, it cannot fully exploit the advantages of neighborhood referencing. On the other hand, setting it to a value close to all reference points would lose the meaning of neighborhood referencing.

At the end, a discriminator $D$ is deployed to enhance $G$ in an adversarial way. $D$'s architecture is revised from the conditional GAN ~\cite{isola2017image}. It takes a (location, fingerprint) pair as input and is trained by four input combinations:

\begin{itemize}
    \item $(u, f^{I2AP}_{u})$, which is considered fake.
    \item $(p_{oth}, f_{u})$, which is considered fake, where $p_{oth}$ is any point other than $u$.
    \item $(u, f_{u})$, which is considered real. During the training stage, the pair $(u,f_u)$  is randomly chosen from $M$.
    \item $(p_{no}, f_{u}+\delta_{no})$, which is not considered real or fake, but is to train $D$ to predict a location, $u^{I2AP}$, as close to $u$ as possible, where $\delta_{no}$ is white noise to disturb the actual fingerprint $f_u$ and $p_{no} \in R^2$ contains simply noise.
\end{itemize}

To cope with the above design, $D$ is slightly different from a typical discriminator. It has to accomplish two tasks: (i) given a (location, fingerprint) pair, it has to decide whether the pair is real or fake, and (ii) given a (random-location, fingerprint) pair, it has to decide the actual location of the fingerprint. The first three input combinations are to train the model to conduct task (i), and the last input combination is to train the model to conduct task (ii). For this purpose, the first part of $D$ is designed to have multiple layers to retrieve common features, and the second part is split into two components, namely classifier $C$ and location predictor $L$, to perform tasks (i) and (ii), respectively. 

The whole model is trained by repeatedly feeding $u$ in $M$ as inputs. During inpainting, $u$ can be any unsurveyed point. We design three loss functions. The first one is for $G$:
\begin{equation}
\begin{aligned}
L_{gen} = \sum_{(u,f_{u})\in M}
\left(
|f^{I2AP}_{u} - f_{u}| +  \log{ D_c({u}, G({u}))}
\right),
\label{loss:generator}
\end{aligned}
\end{equation}
where $D_c$ means the path of classifier $C$ in $D$. Unlike the original cGAN, we add an L1 loss term in Eq.~\ref{loss:generator}. In the image domain, adversarial loss and L1 loss are two antagonism terms. If the weight of the adversarial loss is high, a model would tend to generate a rough image. If the weight of the L1 loss is high, a model would tend to generate a more detailed image. In our experiment, striking a balance between these two terms would perform better.
The second one is for task (i):
\begin{equation}
\begin{aligned}
L_{condition} = \sum_{(u,f_{u})\in M}
\{\log{D_c({u}, f_{u})} \\ + \log{(1 - D_c(p_{oth},f_u))} + 
\log{(1 - D_c({u}, G({u})))}\}.
\label{loss:discriminator}
\end{aligned}
\end{equation}
The above three terms are for the first three input combinations, respectively. 
The third loss function is derived by Euclidean distance and is for task (ii):
\begin{equation}
L_{position} = \sum_{u\in M} |u^{I2AP} - u|.
\end{equation}
Finally, we obtain a weighted total loss function for updating $D$.
\begin{equation}
L_{dis} = \alpha L_{condition} + \beta L_{position}.
\label{loss:I2APtotal_loss_dis}
\end{equation}

\section{Comparisons and Ablation Study}

\begin{figure*}
    \centering
    \includegraphics[width=0.9\textwidth]{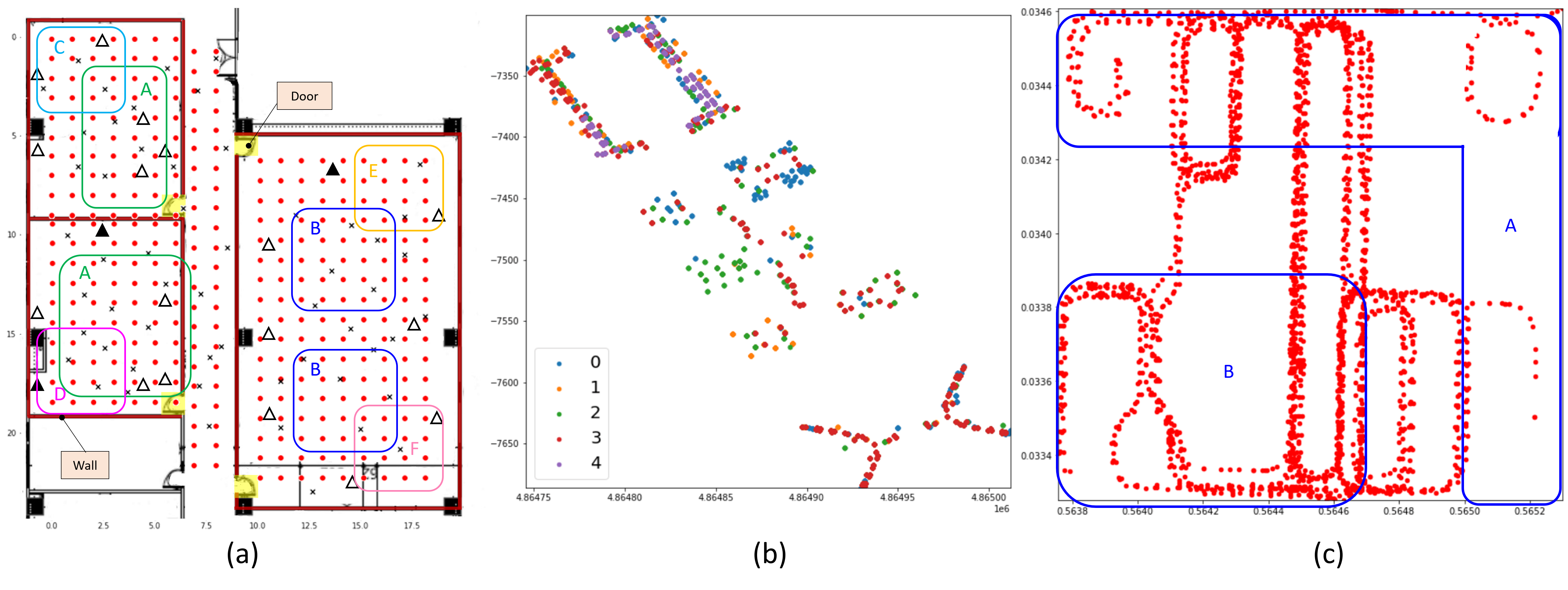}
    \caption{WiFi fingerprint datasets in our validation: (a) The \texttt{campus-multi-fp} dataset has six missing scenarios, A, B, C, D, E and F, each containing one or two unsurvey regions represented by squares. (Red point = grid point; x = non-grid point; solid triangle = 2.4 GHz AP; hollow triangle = 2.4+5 GHz AP) (b) The \texttt{UJIIndoorLoc} dataset has 5 floors, where data points with the same color are on the same floor. (c) The \texttt{MallParking-fp} dataset has two missing scenarios, marked as A and B.}
    \label{fig:Data_sum}
\end{figure*}

We have validated our models on three datasets as shown in Fig.\ref{fig:Data_sum}. \texttt{Campus-multi-fp} was collected from a multi-story campus building in our NYCU Tainan campus. Its second floor has multiple concrete rooms and a corridor with around 20 WiFi APs. Data were collected at 337 grid points and 55 random points, each with around 50 fingerprints. We design six unsurveyed patterns. Patterns A and B are to simulate interior blocks, and patterns C-F are to simulate exterior blocks. Those 55 random points are always regarded unsurveyed. \texttt{UJIIndoorLoc}~\cite{pubilc-dataset} is a public WiFi fingerprint dataset, covering 5 floors, each of size 100 $m^2$. Data were collected at more than 19,000 points from 276 APs, but each point has only one sample. The test set is regarded unsurveyed. The above two are simply point-wise datasets. \texttt{MallParking-fp} was collected in an indoor parking area of a shopping mall in Taipei. To save labor, the fingerprints were collected by using a moving vehicle. Totally 267 APs were observed. Over 2,500 points were collected, each with one fingerprint. Two unsurveyed patterns A and B are designed.

For each designed unsurveyed pattern, missing points will be tested for both inpainting and positioning. We compare our models with two baseline methods, GPR and GPR-GAN~\cite{GPR-GAN}. The training environment is RTX 3090 with an i9-10900k processor and 128GB RAM. The model is implemented using Tensorflow. The details of our models are listed in Table~\ref{table:modeldet}. 

\begin{table} 
\centering
\caption{Model details of IAP and I2AP.}
\setlength{\tabcolsep}{0.1mm}{
\begin{tabular}{|c|c|}
\hline
IAP   $E$:      & \begin{tabular}[c]{@{}c@{}}Dense:128, Dense:192\\     $dim(v_i) = 60$\end{tabular}           \\  \hdashline[1.5pt/5pt]
IAP $D$:        & Dense:128, Dense:64,                                                                        \\
\hline

I2AP $F$:       & \begin{tabular}[c]{@{}c@{}}Dense:256, Dense:128,\\      $k = 4$\textasciitilde$20$, $dim(v_i) = 30,60,120,150$\end{tabular} \\  \hdashline[1.5pt/5pt]
I2AP $G$:     & \begin{tabular}[c]{@{}c@{}}Dense:160, Dense:64, Dense:30\\    \end{tabular}       \\  \hdashline[1.5pt/5pt]
I2AP $D$: & Dense:30, Dense:40, Dense:50, Dense:30                                                        \\ \hline
Optimizer     & \begin{tabular}[c]{@{}c@{}}Adam~\cite{adam}\\ $\beta_1 = 0.9$, $\beta_2 =   0.999$\end{tabular}                                                        \\  \hdashline[1.5pt/5pt]
Learning rate       & $1e^{-4}$                                                                                  \\ \hdashline[1.5pt/5pt]
Training       & Epochs:1000, Batch size:16                                                                                  \\ \hline
\end{tabular}}
\label{table:modeldet}
\end{table}

\subsection{Performance Comparisons}

\begin{figure} 
    \centering
    \includegraphics[width=0.45\textwidth]{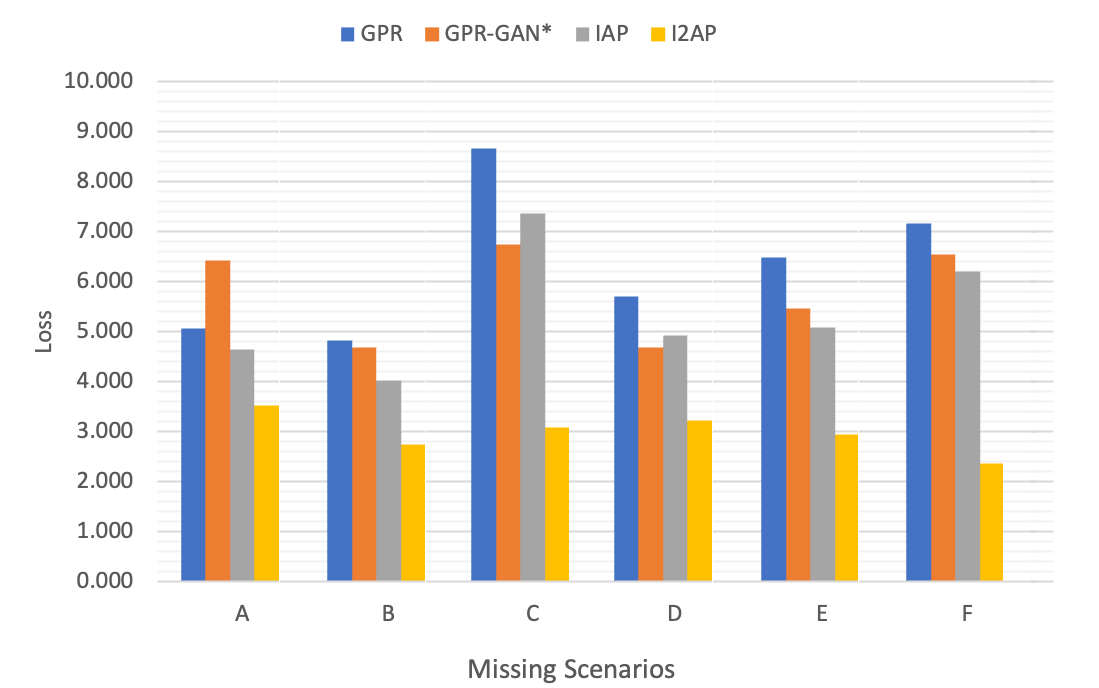}
    \caption{Comparisons of inpainting errors on \texttt{Campus-multi-fp} (L1 loss, $k=4$, and $dim(v_i) = 30$).}
    \label{fig:loss_table}
\end{figure}

\textbf{Inpainting Errors on \texttt{Campus-multi-fp}:} Figure~\ref{fig:loss_table} shows the L1 loss of different inpainting solutions. We can observe that GPR is highly sensitive to the locations of the inpainted areas. It incurs less loss for interior areas i.e., patterns A and B, but incurs larger loss for exterior areas, i.e., patterns, C-F. This is because GPR is a kernel-based method, so its uncertainty increases as the unsurveyed region is out of the surveyed area. Although IAP performs better than GPR on all patterns, it still suffers from poorer performance on the exterior block patterns. One major reason is that IAP's inputs are computed by GPR. Once GPR cannot provide a good start for IAP, the outcome will degrade. GPR-GAN$^*$ (reproduced from GPR-GAN by ourselves) suffers from a similar problem to IAP. It is not so stable since it also relies on GPR's input. Moreover, without sufficient training data, it is hard to train the GAN-based model. I2AP works much better and shows good stability no matter in the interior or exterior missing scenarios. Via exploiting intra-AP and inter-AP correlations at the same time, I2AP shows lower L1 loss than the previous solutions. This validates that both correlations are important for the inpainting task. I2AP overcame the drawback of GPR-based solutions and performed the best in all scenarios.

\textbf{Inpainting Errors on \texttt{MallParking-fp}:}
The dataset was collected from a slow-moving vehicle. We treat each collected data as ground truth.
Table~\ref{table:C-I2AP_compare_101} shows the L1 loss of GPR and I2AP. Overall, since each point in this dataset only has one sample, we observe that the L1 loss is higher compared to those of \texttt{Campus-multi-fp}. Our results indicate that I2AP outperforms GPR. Furthermore, we can observe the adaptability of I2AP when dealing with exterior unsurveyed areas. Despite the challenges posed by sparser data points and the presence of only one fingerprint for each data point, I2AP demonstrates superior performance compared to GPR, highlighting the robustness of I2AP in handling variations in data.

\begin{table} 
\centering
\caption{Comparisons of inpainting errors on \texttt{MallParking-fp} (L1 loss, $k=4$, and $dim(v_i) = 160$).}
\begin{tabular}{|l|l|l|l|}
\hline
L1 Loss            & GPR   & I2AP    \\ \hline
Pattern A & 15.76 & \textbf{13.18}  \\ \hline
Pattern B & 13.93 & \textbf{13.91}  \\ \hline
\end{tabular}
\label{table:C-I2AP_compare_101}
\end{table}

\textbf{Inpainting Errors on \texttt{UJIIndoorLoc}:}
In this public dataset, rather than repartitioning the data into training and testing sets on our own, we used its own split sets for testing. As illustrated in Figure \ref{fig:Data_sum}(b), the whole areas are covered and there are no unsurveyed regions. So we regard the training set as $M$ and the test set as unsurveyed. The results are in Table \ref{error-UJI}, which shows that I2AP outperforms GPR in most cases. We attribute the closer performance of the two models to the fact that the unsurveyed points are all surrounded by surveyed points in this dataset, making the inpainting task much easier.

\begin{table} 
\centering
\caption{Comparisons of inpainting errors on \texttt{UJIIndoorLoc} (L1 loss, $k=20$, and $dim(v_i) = 160$).}
\begin{tabular}{|c|c|c|c|}
\hline
Floor & GPR     & I2AP                      \\ \hline
0     & 13.0292 & \textbf{9.6336}           \\ \hline
1     & 10.1400 & \textbf{9.5474}           \\ \hline
2     & \textbf{9.2411}  & 9.6552           \\ \hline
3     & 11.3369 & \textbf{10.2124}          \\ \hline
4     & 13.3063 & \textbf{9.1633}           \\ \hline
\end{tabular}
\label{error-UJI}
\end{table}

\begin{table*}[htp]
\centering
\caption{CEP68 and CEP95 for KNN positioning on \texttt{Campus-multi-fp}: (a) grid-point test and (b)non-grid-point test.}
\begin{tabular}{|cccccc|c|cccccc|}
\cline{1-6} \cline{8-13}
\multicolumn{6}{|c|}{KNN CEP68(m)}                                                                                                                              &  & \multicolumn{6}{c|}{KNN CEP68(m)}                                                                                                                                            \\ \cline{1-6} \cline{8-13} 
\multicolumn{1}{|c|}{}  & \multicolumn{1}{c|}{Baseline} & \multicolumn{1}{c|}{GPR}  & \multicolumn{1}{c|}{GPR-GAN*} & \multicolumn{1}{c|}{IAP}  & I2AP          &  & \multicolumn{1}{c|}{}  & \multicolumn{1}{c|}{Baseline}      & \multicolumn{1}{c|}{GPR}  & \multicolumn{1}{c|}{GPR-GAN*} & \multicolumn{1}{c|}{IAP}           & I2AP          \\ \cline{1-6} \cline{8-13} 
\multicolumn{1}{|c|}{A} & \multicolumn{1}{c|}{2.72}     & \multicolumn{1}{c|}{2.68} & \multicolumn{1}{c|}{2.73}     & \multicolumn{1}{c|}{2.63} & \textbf{2.62} &  & \multicolumn{1}{c|}{A} & \multicolumn{1}{c|}{2.67}          & \multicolumn{1}{c|}{2.75} & \multicolumn{1}{c|}{2.68}     & \multicolumn{1}{c|}{2.84}          & \textbf{2.59} \\ \cline{1-6} \cline{8-13} 
\multicolumn{1}{|c|}{B} & \multicolumn{1}{c|}{2.59}     & \multicolumn{1}{c|}{2.52} & \multicolumn{1}{c|}{2.61}     & \multicolumn{1}{c|}{2.39} & \textbf{2.05} &  & \multicolumn{1}{c|}{B} & \multicolumn{1}{c|}{2.92}          & \multicolumn{1}{c|}{2.94} & \multicolumn{1}{c|}{2.99}     & \multicolumn{1}{c|}{\textbf{2.91}} & 3.01          \\ \cline{1-6} \cline{8-13} 
\multicolumn{1}{|c|}{C} & \multicolumn{1}{c|}{3.58}     & \multicolumn{1}{c|}{3.58} & \multicolumn{1}{c|}{3.58}     & \multicolumn{1}{c|}{3.56} & \textbf{2.82} &  & \multicolumn{1}{c|}{C} & \multicolumn{1}{c|}{\textbf{2.62}} & \multicolumn{1}{c|}{2.73} & \multicolumn{1}{c|}{2.69}     & \multicolumn{1}{c|}{2.86}          & 2.69          \\ \cline{1-6} \cline{8-13} 
\multicolumn{1}{|c|}{D} & \multicolumn{1}{c|}{4.15}     & \multicolumn{1}{c|}{4.15} & \multicolumn{1}{c|}{4.15}     & \multicolumn{1}{c|}{3.99} & \textbf{3.75} &  & \multicolumn{1}{c|}{D} & \multicolumn{1}{c|}{2.70}          & \multicolumn{1}{c|}{2.80} & \multicolumn{1}{c|}{2.92}     & \multicolumn{1}{c|}{3.01}          & \textbf{2.70} \\ \cline{1-6} \cline{8-13} 
\multicolumn{1}{|c|}{E} & \multicolumn{1}{c|}{4.44}     & \multicolumn{1}{c|}{4.44} & \multicolumn{1}{c|}{4.17}     & \multicolumn{1}{c|}{4.17} & \textbf{3.08} &  & \multicolumn{1}{c|}{E} & \multicolumn{1}{c|}{2.80}          & \multicolumn{1}{c|}{2.78} & \multicolumn{1}{c|}{3.09}     & \multicolumn{1}{c|}{3.09}          & \textbf{2.56} \\ \cline{1-6} \cline{8-13} 
\multicolumn{1}{|c|}{F} & \multicolumn{1}{c|}{4.30}     & \multicolumn{1}{c|}{4.30} & \multicolumn{1}{c|}{4.32}     & \multicolumn{1}{c|}{4.3}  & \textbf{2.93} &  & \multicolumn{1}{c|}{F} & \multicolumn{1}{c|}{\textbf{2.44}} & \multicolumn{1}{c|}{2.46} & \multicolumn{1}{c|}{2.68}     & \multicolumn{1}{c|}{2.88}          & 2.48          \\ \cline{1-6} \cline{8-13} 
\multicolumn{6}{|c|}{KNN CEP95(m)}                                                                                                                              &  & \multicolumn{6}{c|}{KNN CEP95(m)}                                                                                                                                            \\ \cline{1-6} \cline{8-13} 
\multicolumn{1}{|c|}{A} & \multicolumn{1}{c|}{4.47}     & \multicolumn{1}{c|}{4.58} & \multicolumn{1}{c|}{4.47}     & \multicolumn{1}{c|}{4.48} & \textbf{4.17} &  & \multicolumn{1}{c|}{A} & \multicolumn{1}{c|}{4.65}          & \multicolumn{1}{c|}{4.74} & \multicolumn{1}{c|}{4.74}     & \multicolumn{1}{c|}{4.40}          & \textbf{4.13} \\ \cline{1-6} \cline{8-13} 
\multicolumn{1}{|c|}{B} & \multicolumn{1}{c|}{3.92}     & \multicolumn{1}{c|}{3.80} & \multicolumn{1}{c|}{3.94}     & \multicolumn{1}{c|}{3.56} & \textbf{3.15} &  & \multicolumn{1}{c|}{B} & \multicolumn{1}{c|}{4.81}          & \multicolumn{1}{c|}{4.79} & \multicolumn{1}{c|}{4.79}     & \multicolumn{1}{c|}{\textbf{4.65}} & 4.77          \\ \cline{1-6} \cline{8-13} 
\multicolumn{1}{|c|}{C} & \multicolumn{1}{c|}{5.78}     & \multicolumn{1}{c|}{5.78} & \multicolumn{1}{c|}{5.78}     & \multicolumn{1}{c|}{5.28} & \textbf{4.51} &  & \multicolumn{1}{c|}{C} & \multicolumn{1}{c|}{\textbf{4.14}} & \multicolumn{1}{c|}{4.27} & \multicolumn{1}{c|}{4.65}     & \multicolumn{1}{c|}{4.56}          & 4.65          \\ \cline{1-6} \cline{8-13} 
\multicolumn{1}{|c|}{D} & \multicolumn{1}{c|}{5.94}     & \multicolumn{1}{c|}{5.94} & \multicolumn{1}{c|}{5.94}     & \multicolumn{1}{c|}{5.63} & \textbf{5.80} &  & \multicolumn{1}{c|}{D} & \multicolumn{1}{c|}{4.65}          & \multicolumn{1}{c|}{4.65} & \multicolumn{1}{c|}{4.65}     & \multicolumn{1}{c|}{5.01}          & \textbf{4.65} \\ \cline{1-6} \cline{8-13} 
\multicolumn{1}{|c|}{E} & \multicolumn{1}{c|}{5.72}     & \multicolumn{1}{c|}{5.72} & \multicolumn{1}{c|}{5.60}     & \multicolumn{1}{c|}{5.60} & \textbf{4.81} &  & \multicolumn{1}{c|}{E} & \multicolumn{1}{c|}{4.78}          & \multicolumn{1}{c|}{4.78} & \multicolumn{1}{c|}{5.27}     & \multicolumn{1}{c|}{5.27}          & \textbf{4.65} \\ \cline{1-6} \cline{8-13} 
\multicolumn{1}{|c|}{F} & \multicolumn{1}{c|}{5.11}     & \multicolumn{1}{c|}{5.16} & \multicolumn{1}{c|}{5.11}     & \multicolumn{1}{c|}{5.15} & \textbf{3.94} &  & \multicolumn{1}{c|}{F} & \multicolumn{1}{c|}{\textbf{4.65}} & \multicolumn{1}{c|}{4.65} & \multicolumn{1}{c|}{5.10}     & \multicolumn{1}{c|}{4.97}          & 4.85          \\ \cline{1-6} \cline{8-13} 

\end{tabular}
\label{table:cep}
\end{table*}

\textbf{Positioning Accuracy:} Once we inpaint fingerprint data for unsurveyed regions, we can use the new data for indoor positioning. Therefore, we implement a simple KNN fingerprint-based positioning method \cite{wifi_pos_3}.
With this downstream task, we further evaluate the contribution of the inpainted data. The performance metric used here is the Circular Error Probability (CEP)~\cite{CPE_acc}. A lower CEP value stands for higher positioning accuracy. For example, when CEP-$q$ = $p$ meters, it means that positioning errors for the most $q\%$ accurate positioning results are less than or equal to $p$ meters. Two typical measurements, CEP68 and CEP95, are used here, which represent median and high confidence positioning errors, respectively. Table~\ref{table:cep} shows the CEP values when using grid and non-grid (i.e., random) points for positioning. We take the \texttt{Campus-multi-fp} dataset for evaluation (refer to Fig. \ref{fig:Data_sum}(a)), which has 55 non-grid points. The difference here is that these 55 non-grid points may fall in both surveyed and unsurveyed areas. Intuitively, those inpainted points falling in the surveyed regions may make less contributions to positioning. As shown in Table~\ref{table:cep}, I2AP is very competitive no matter the test dataset is composed of grid or non-grid points.

\subsection{Ablation Study}
In Table~\ref{table:I2API_compare}, we compare I2AP with two variants, I2AP$^{-}$ (without module $L$) and I2AP$^{--}$ (without discriminator $D$). Basically, we can see that each enhancement does make a clear contribution. In terms of the positioning error for non-grid points, as we have pointed out earlier, inpainting the unsurveyed regions may not play a key role for positioning. 

\begin{table}[ht]
\centering
\caption{Ablation study on \texttt{Campus-multi-fp}, Pattern B. L1 loss and positioning error are compared.}
\setlength{\tabcolsep}{0.5mm}{
\begin{tabular}{|l|c|c|c|c|c|}
\hline
        & L1 loss       & \begin{tabular}[c]{@{}c@{}}grid\\ CEP68\end{tabular} & \begin{tabular}[c]{@{}c@{}}grid\\ CEP95\end{tabular} & \begin{tabular}[c]{@{}c@{}}non-grid\\ CEP68\end{tabular} & \begin{tabular}[c]{@{}c@{}}non-grid \\ CEP95\end{tabular} \\ \hline
I2AP$^{--}$  & 2.900        & 2.114        & 3.637                & 2.853           & \textbf{4.654}                              \\ \hline
I2AP$^{-}$ & 2.8856         & \textbf{2.0300}                                                 & 3.499                                                 & \textbf{2.795}                                        & \textbf{4.654}                                                \\ \hline
I2AP    & \textbf{2.7271}          & 2.056                                     & \textbf{3.155}                                       & 3.015                                                 & 4.772                                          \\ \hline
\end{tabular}
}
\label{table:I2API_compare}
\end{table}

\section{Conclusions}
Indoor localization is a critical problem for intelligent business and robotics. Previous image and video inpainting models are mostly trained by a large amount of data. WiFi inpainting, however, can hardly rely on past experiences or data because a fingerprint map is highly dependent on the indoor physical signal propagation model, environment, and APs' deployment. To conquer these challenges, we have proposed two models capable of adapting to various WiFi fingerprint scenarios and thoroughly exploring spatial relationships. Different missing patterns have been tested, and the inpainted fingerprints do help improve positioning accuracy. As to future directions, expanding the models by incorporating a larger volume of crowdsourced data or delving deeper into understanding the effects of environmental factors and the property of WiFi signal attenuation would deserve further exploration.

\bibliographystyle{IEEEtran}
\bibliography{ref}


\end{document}